\documentclass{nature_fig}

\usepackage{sansmath}
\usepackage{amssymb}

\usepackage{epsfig}
\usepackage{graphicx}
\usepackage{upgreek}
\usepackage{xcolor}
\usepackage{cite}
\usepackage{caption}

\title{Probing supernovae and kicks in post-supernova binaries}

\author{C.~Larsen$^1$, H.~C.~G.~Larsen$^1$, C.~C.~Pedersen$^1$, P.~N.~Thomsen$^1$, J.~Tøffner-Clausen$^1$, T.~M.~Tauris$^{1*}$
}

\newcommand{\aap}{Astronomy and Astrophysics}
\newcommand{\apj}{The Astrophysical Journal}
\newcommand{\apjl}{The Astrophysical Journal Letters}

\newcommand{\mnras}{Monthly Notices of the Royal Astronomical Society}

\newcommand{\prl}{Physical Review Letters}

\begin{document}

\maketitle

\begin{affiliations}
 \item Department of Materials and Production, Aalborg University, Denmark
 \item[*] e-mail: tauris@mp.aau.dk
\end{affiliations}

\begin{abstract}
Knowledge of the formation of neutron stars (NSs) in supernova (SN) explosions is of fundamental importance in wide areas of contemporary astrophysics: X-ray binaries, magnetars, radio pulsars, and, not least, double NS systems which merge and become gravitational wave sources. A recent study by Richardson~et~al.\cite{rpe+23} reported that the NS in the Be-star/X-ray binary SGR~0755$-$2933 (CPD~$-$29~2176)\cite{bdg+16,dstj21} descended from an ultra-stripped SN. Using the same observational data as Richardson~et~al.\cite{rpe+23}, however, we find that the majority of progenitor solutions for SGR~0755$-$2933 are of normal Type~Ib/c SNe, which allows for up to several solar masses of material to be ejected in the SN event.
To correctly probe the SN explosion physics and inferring pre-SN conditions in a binary system, a full kinematic analysis based on post-SN data is always needed.
\end{abstract}

Richardson~et~al.\cite{rpe+23} investigated the progenitor evolution of the binary system that created SGR~0755$-$2933\cite{bdg+16,dstj21}. Their key assumptions that i) a near-circular post-SN system requires very little ejecta mass, and, as a result of that, ii) an ultra-stripped SN must have been at work (according to their claim such SNe produce little to no kick) are, unfortunately, not correct.

The fact that a given post-SN system is near-circular --- assuming here tidal interactions being insignificant for such a wide-orbit (59.7~days) system like SGR~0755$-$2933 --- will indeed constrain the family of SN solutions at work. These solutions, however, are based on a combination of the momentum kick (magnitude and direction) imparted onto the newborn NS and the amount of SN ejecta mass. An analysis of the mechanics equations\cite{hil83} applied to solve for the kinematic solutions of a post-SN binary with component masses, eccentricity and orbital period resembling those of SGR~0755$-$2933 reveals results that are in stark contrast to the findings of Richardson~et~al.\cite{rpe+23}
In Figure~1, we plot the post-SN eccentricity of binaries with an assumed NS mass of $1.44\;M_\odot$ and a companion star mass of $18\;M_\odot$ (adopting the values for SGR~0755$-$2933 applied by Richardson~et~al.\cite{rpe+23}), but assuming here an exploding helium star mass of $3.6~M_\odot$. The important point to notice from this plot is that near-circular (i.e. eccentricity, $e<0.1$) post-SN binaries may well be produced even if i) the exploding star ejects about $2.0\;M_\odot$ of material (released gravitational binding energy reduces the newborn NS mass by another $\sim 0.17\;M_\odot$), and ii) the kick magnitude is relatively large. The outcome depends crucially on the direction of the kick (defined by the two kick angles, $\theta$ and $\phi$; see also Extended Data~Fig.\,1)\cite{hil83,tkf+17,tv23}. More specifically, the plot shows a valley of solutions of low eccentricity even for quite substantial kick magnitudes up to $w/v_{\rm rel} \sim 2.0$ (roughly corresponding to kicks of $w\simeq 300\;{\rm km\,s}^{-1}$, given the typical pre-SN orbital period of about 55~days required to reproduce SGR~0755$-$2933 after the SN). 
Figure~2 shows a histogram of the inferred mass of the exploding star that produced the NS in SGR~0755$-$2933 based on our Monte Carlo simulations (see Methods) of 500,000 SN explosions that successfully reproduced the observed properties of the SGR~0755$-$2933 system: orbital period of $P_{\rm orb}=59.7~{\rm days}$ (within an acceptable range of $\pm 3\%$) and a low eccentricity, $e=0.00-0.12$.

Despite that both large kicks and large amounts of ejecta mass on their own may give rise to high post-SN eccentricities, it is no surprise that solutions for a near-circular orbit exist because these two effects may cancel each other out, depending on the kick direction.
In contrast to the conclusion reached by Richardson~et~al.\cite{rpe+23}, that the exploding star must have been ultra-stripped prior to the SN, we find that the far majority ($92.4\%$) of the kinematic solutions for SGR~0755$-$2933 had an exploding star with an ejecta mass exceeding $0.2\;M_\odot$ (in average close to $1.5\;M_\odot$), which means a ``normal'' Type~Ib/c SN rather than an ultra-stripped SN\cite{tlp15}. If we take a less conservative upper limit for the mass of the exploding star, and allow masses above $5.0\;M_\odot$, then the average amount of ejecta mass will be even larger. 
Our simulations reveal a mean kick of $25\;{\rm km\,s}^{-1}$ (although rare solutions also occur for kicks up to $\sim 300\;{\rm km\,s}^{-1}$; the median and equal-tailed 90\% credible interval are $11\;{\rm km\,s}^{-1}$ and $[1.2-99]_{90}\;{\rm km\,s}^{-1}$, respectively, see Extended Data Fig.\,2) and that these kicks, combined with the instantaneous mass loss, result in systemic recoil velocities of the system with a median value of $11\;{\rm km\,s}^{-1}$ ($[2.5-23]_{90}\;{\rm km\,s}^{-1}$, Extended Data Fig.\,3), in accordance with the peculiar velocity estimated by Richardson~et~al.\cite{rpe+23} of $15.3\pm3.0\;{\rm km\,s}^{-1}$.

To strip the remaining envelope of a collapsing helium star all the way down to $\lesssim 0.2\;M_\odot$ via mass transfer prior to the SN, as proposed by Richardson~et~al.\cite{rpe+23}, is not possible with a non-degenerate star accretor\cite{ywl10,ljr+21}. Only in systems where the stripping of the helium star is caused by a compact object it is possible to peel off the outer layers via mass transfer (so-called Case~BB Roche-lobe overflow) all the way down to an almost naked metal core before core collapse\cite{tlp15}.   
Because the NS in SGR~0755$-$2933 has an OB-star companion it was created in the first SN of that system. It is therefore expected also from a binary stellar evolution point of view\cite{tv23} that it was produced by a normal Type~Ib/c SN. Ultra-stripped SNe\cite{dsm+13,dko+18}, on the other hand, are only expected as the second SN in a tight binary, often producing double NS\cite{tlm+13,mmt+17,tkf+17,jtcf21} or black hole--NS binaries\cite{aaa+17c}; SGR~0755$-$2933 is not an exception.

To probe the SN explosion physics and inferring pre-SN conditions based on post-SN data of a binary system, we conclude that care should be taken to include a full kinematic analysis.
Finally, we remark that the proposed future evolution scenario of Richardson~et~al.\cite{rpe+23}, namely that SGR~0755$-$2933 may become a double NS system, is highly questionable. The reason is that the present NS is most likely unable to eject the common envelope once its $18\;M_\odot$ Be-star companion expands as a giant star and captures the NS. Only very wide-orbit ($>1\;{\rm yr}$) Be-star/X-ray binaries are expected to have small enough envelope binding energies\cite{tkf+17} that will allow for common envelope ejection during NS in-spiral, thereby paving the way for a tight system that undergoes a second (this time most likely ultra-stripped) SN.

\noindent
{\bf Methods}\\
Our Monte Carlo simulations follow the method of ref.\cite{tkf+17} and were calculated in two independent ways (for a double check) using Python and FORTRAN.
We applied the parameter values adopted by Richardson~et~al.\cite{rpe+23}: $1.44\;M_\odot$ for the NS, $18\;M_\odot$ for its companion, orbital eccentricity of $e\le 0.12$ and orbital period of 59.7~days. Given the wide orbit of the system, it is safe to neglect any orbital evolution since its formation. 
We simulated SNe over a 5-dimensional phase~space whose parameters are: pre-SN orbital period, mass of the exploding star, magnitude of the NS kick velocity, and the two kick direction angles, $\theta$ and $\phi$. 
We assumed flat non-informative prior distributions for all these parameters except for $\theta$ where a random (isotropic) kick direction leads to a prior distribution of $\sin(\theta)$.
Using Monte Carlo methods, we repeatedly select a set of values for the above-mentioned five parameters and solve in each trial for the post-SN parameters.
These can be compared with the values of the SGR~0755$-$2933 system and are selected as valid solutions if within a chosen error margin of 3\% in orbital period and an eccentricity, $e\le 0.12$.

\noindent
{\bf Data availability}
The simulated solution data shown in Fig.~2 and Extended Data Fig.~2 are available from the corresponding author upon request. 

\noindent
{\bf Code availability}
The Python scripts for producing all figures are available from the corresponding author upon request.

\clearpage
\begin{figure*}[h]
\includegraphics[width=1.0\textwidth]{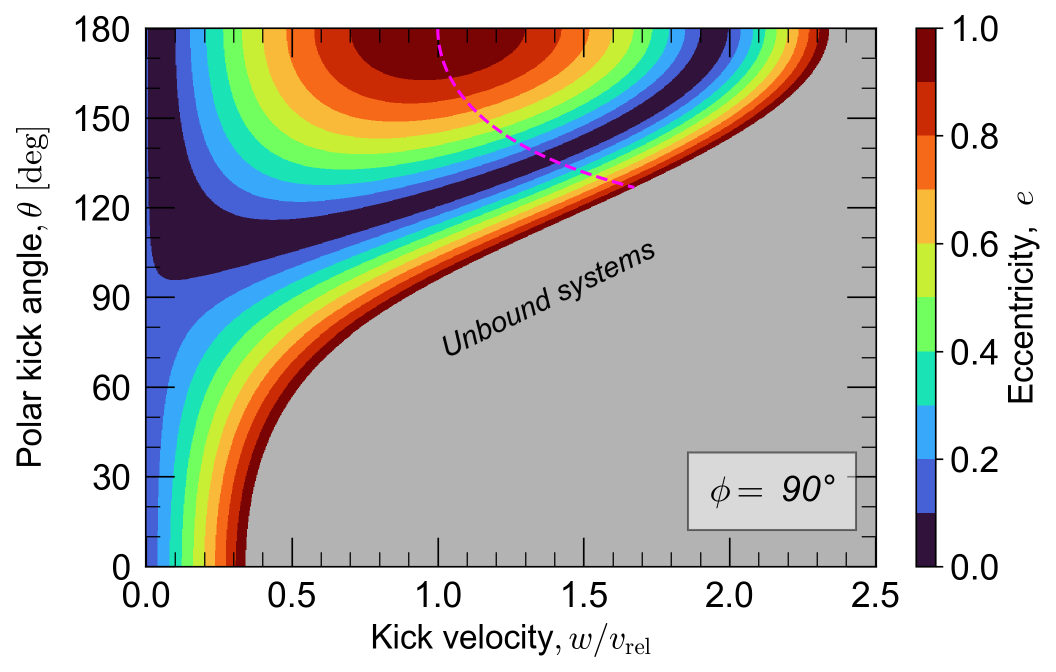}
\caption*{{\bf Figure\,1}\;\;
\textbf{Eccentricity as function of SN kick velocity and polar kick angle for a post-SN system resembling SGR~0755$-$2933.}
The eccentricity is color-coded between 0 and 1, and the SN kick velocity magnitude, $w$, is in units of the pre-SN relative orbital velocity, $v_{\rm rel}$. The plot is based on analytical calculations assuming as an example here a secondary kick angle, $\phi = 90^\circ$, a mass of the exploding progenitor (helium) star of $3.6\;M_\odot$, a resulting NS mass of $1.44\;M_\odot$, and a companion star mass of $18\;M_\odot$. 
The grey area indicates the region where post-SN systems are unbound.
The dashed line indicates the boundary where the critical polar kick angle, $\theta_{\rm retro} = \cos^{-1}(-v_{\rm rel}/w)$, separates post-SN prograde (left) and retrograde orbits (right).
\label{fig:eccentricity}
}
\end{figure*}

\clearpage
\begin{figure*}[h]
\hspace*{1.0cm}
\includegraphics[width=0.8\textwidth]{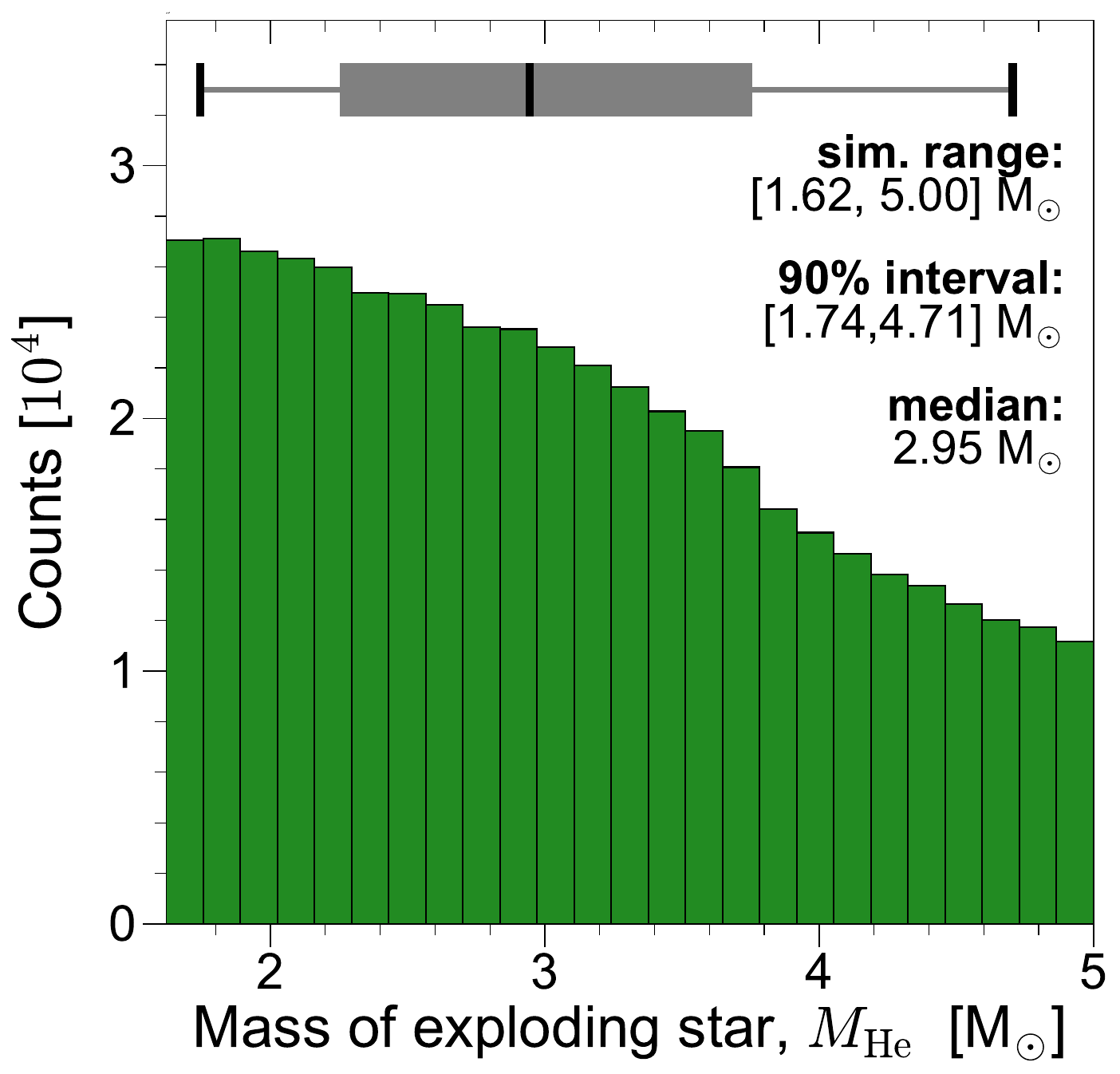}
\caption*{{\bf Figure\,2}\;\;
\textbf{Distribution of count of the mass of the exploding star that produced the NS in SGR~0755$-$2933.}
The plot is based on Monte Carlo simulations of 500,000 SN explosions that successfully reproduced the observed properties of the SGR~0755$-$2933 system: orbital period of $P_{\rm orb}=59.7~{\rm days}$ (within a range of $\pm 3\%$) and eccentricity, $e=0.00-0.12$. The median value of the distribution and equal-tailed 50\% (grey) and 90\% (full range) credible intervals are plotted on top.
\label{fig:progenitor_mass}
}
\end{figure*}


\clearpage
\noindent
{\bf References}


\begin{addendum}
 \item We thank Johan Fynbo for stimulating discussions and BSc project examination.
 
\item[Author contributions] All authors contributed equally. C.L., H.C.G.L., C.C.P., P.N.T., J.T. prepared the code, derived analytical equations and investigated background material. T.M.T. outlined the project, made the initial computations, and drafted the manuscript. All authors discussed the results.
 
\item[Competing interests] The authors declare no competing interests.

\item[Correspondence and request for materials] Correspondence and requests for materials
should be addressed to T.M.T.~(email: tauris@mp.aau.dk).
\end{addendum}

\newpage
\section*{{\Large EXTENDED DATA}}
\vspace{2.0cm}
\begin{figure*}[h]
\includegraphics[width=1.0\textwidth]{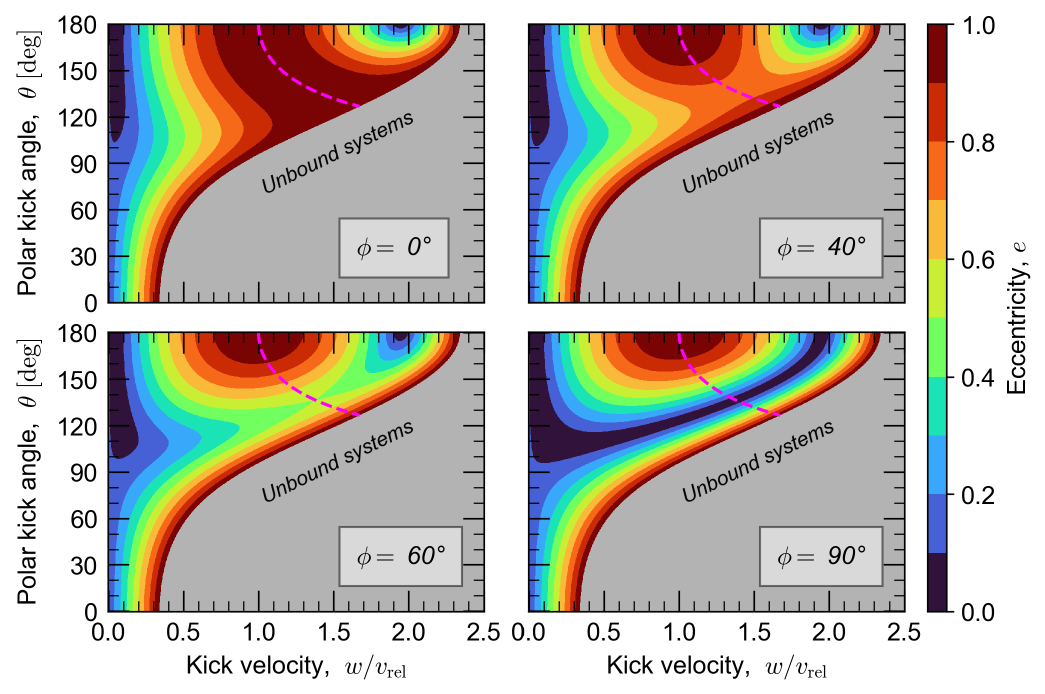}

\caption*{{\bf Extended Data Fig.\,1}\;\;
\textbf{Eccentricity as function of SN kick velocity and polar kick angle, using four different secondary kick angles, for a post-SN system resembling SGR~0755$-$2933.}
The eccentricity is color-coded between 0 and 1, the SN kick velocity magnitude, $w$, is in units of the pre-SN relative orbital velocity, $v_{\rm rel}$, and the secondary kick angles are $\phi = 0^\circ$ (top left), $\phi = 40^\circ$ (top right), $\phi = 60^\circ$ (bottom left), and $\phi = 90^\circ$ (bottom right). The plots are based on analytical calculations assuming a mass of the exploding progenitor (helium) star of $3.6\;M_\odot$, a resulting NS mass of $1.44\;M_\odot$, and a companion star mass of $18\;M_\odot$. 
The grey area indicates the region where post-SN systems are unbound.
The dashed line indicates the boundary where the critical polar kick angle, $\theta_{\rm retro} = \cos^{-1}(-v_{\rm rel}/w)$, separates post-SN prograde (left) and retrograde orbits (right).
}
\end{figure*}

\begin{figure*}[h]
\vspace*{-1.0cm}
\hspace*{0.4cm}
\includegraphics[width=0.37\textwidth]{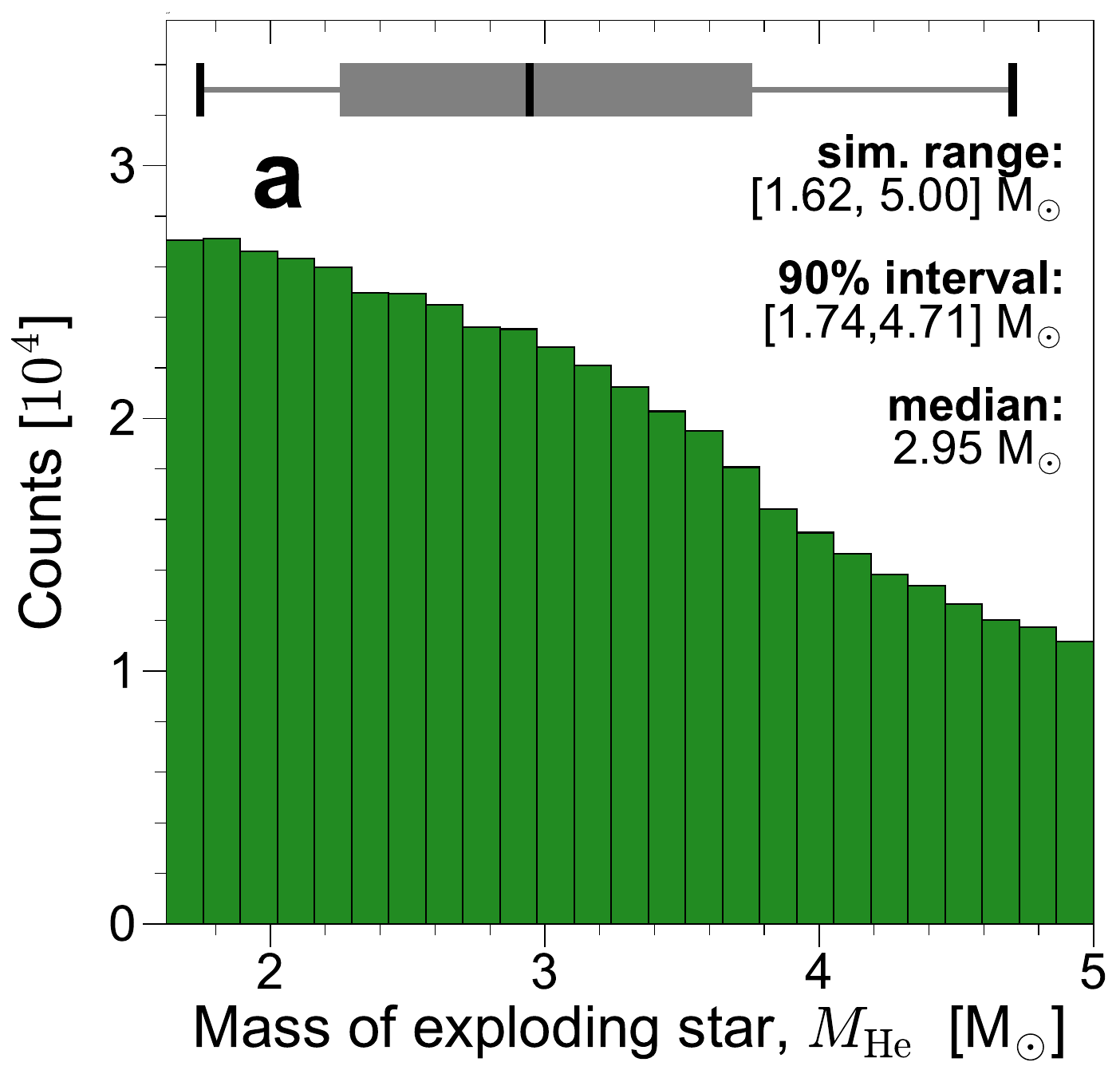}
\hspace*{0.9cm}
\includegraphics[width=0.375\textwidth]{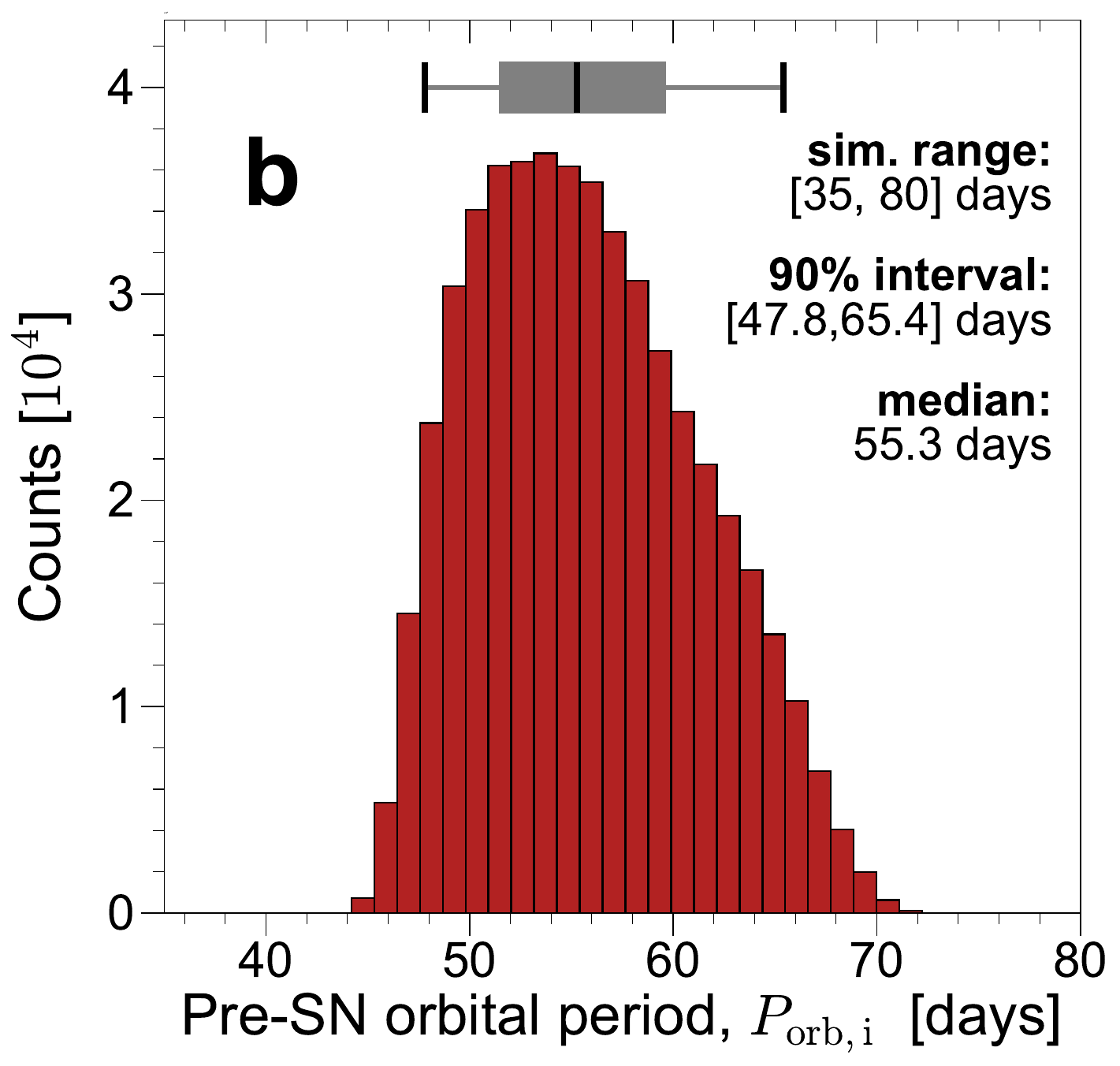}

\vspace*{0.2cm}
\hspace*{0.52cm}
\includegraphics[width=0.36\textwidth]{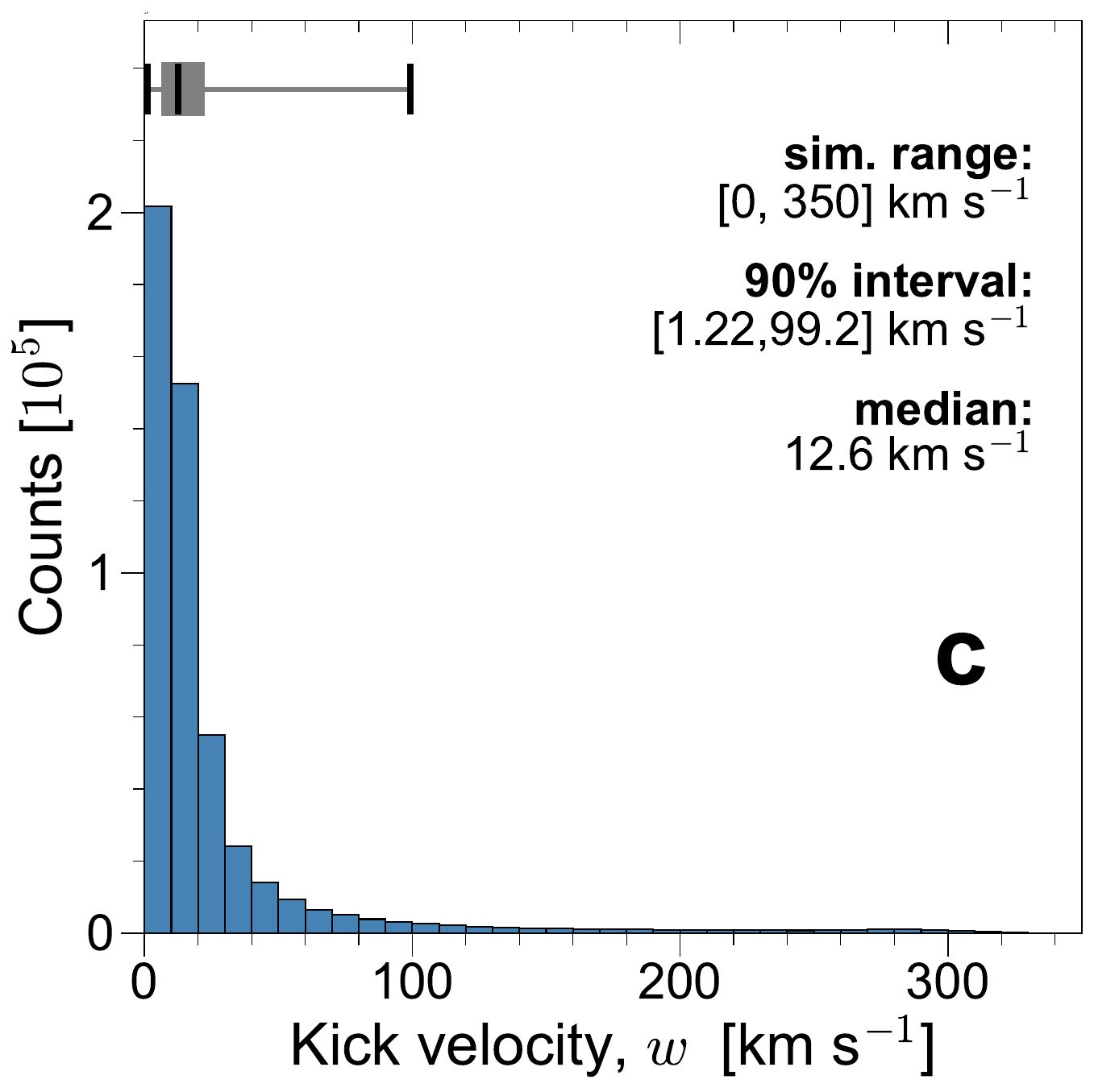}
\hspace*{1.0cm}
\vspace*{-0.4cm}
\includegraphics[width=0.36\textwidth]{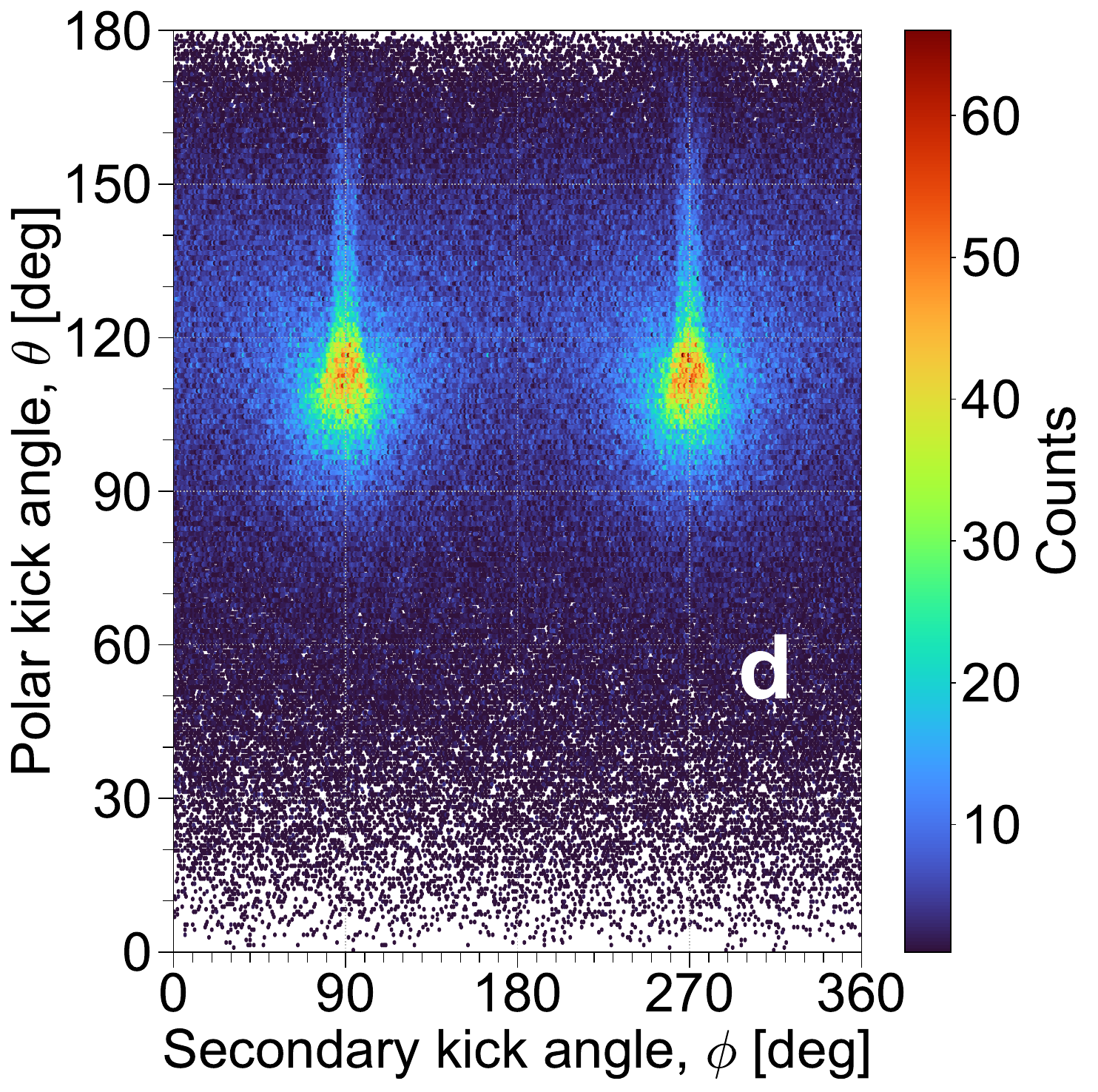}

\vspace*{0.6cm}
\hspace*{0.40cm}
\includegraphics[width=0.382\textwidth]{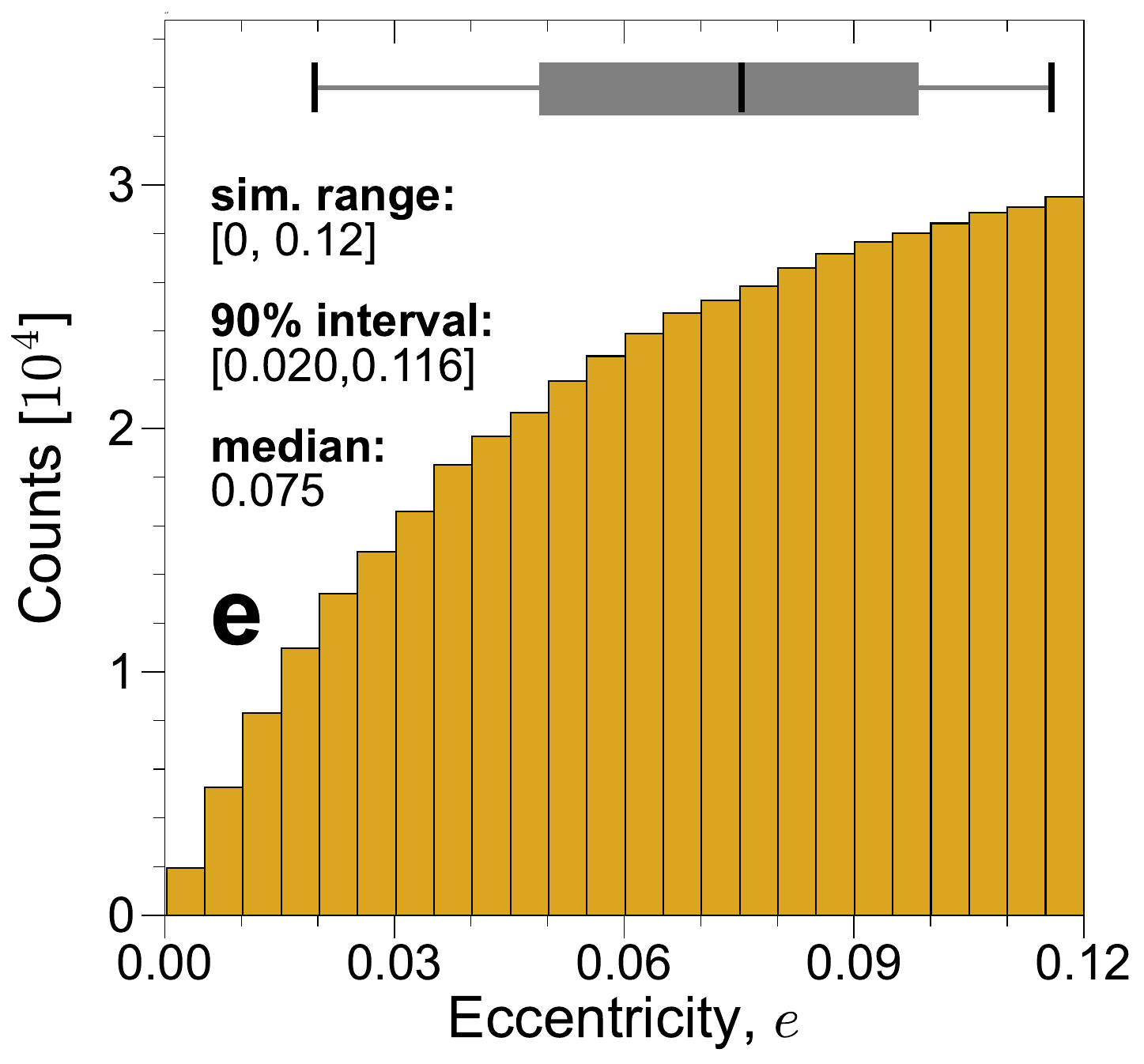}
\hspace*{0.7cm}
\includegraphics[width=0.375\textwidth]{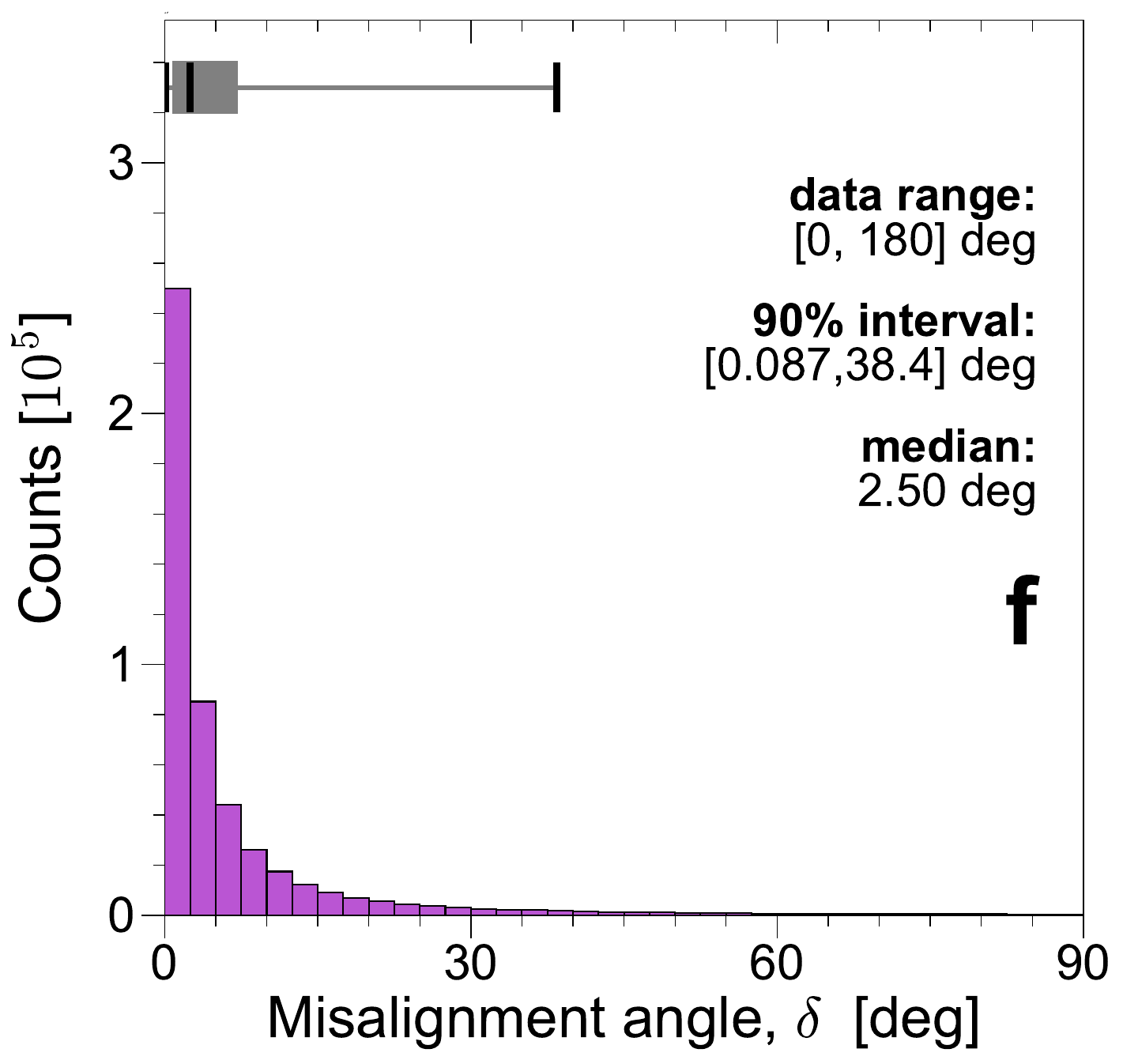}

\caption*{{\bf Extended Data Fig.\,2}\;\;
\textbf{Constraints on the formation of the SGR~0755$-$2933 system.} 
The plots are based on Monte Carlo simulations of 500,000 SN explosions that successfully reproduced the observed parameters of SGR~0755$-$2933: post-SN orbital period of $P_{\rm orb}=59.7~{\rm days}$ (within an acceptable range of $\pm 3\%$) and an eccentricity, $e=0.00-0.12$. The six panels ({\bf a}--{\bf f}) display distributions of: pre-SN mass of exploding star ({\bf a}), pre-SN orbital period ({\bf b}), magnitude of SN kick velocity ({\bf c}), direction of SN kick velocity ({\bf d}), post-SN eccentricity ({\bf e}), and post-SN misalignment angle ({\bf f}, i.e. tilt of the orbital plane due to the SN kick). The median and equal-tailed 50\% (grey) and 90\% (full range) credible intervals are plotted. 
}
\end{figure*}

\begin{figure*}[h]
\includegraphics[width=0.7\textwidth]{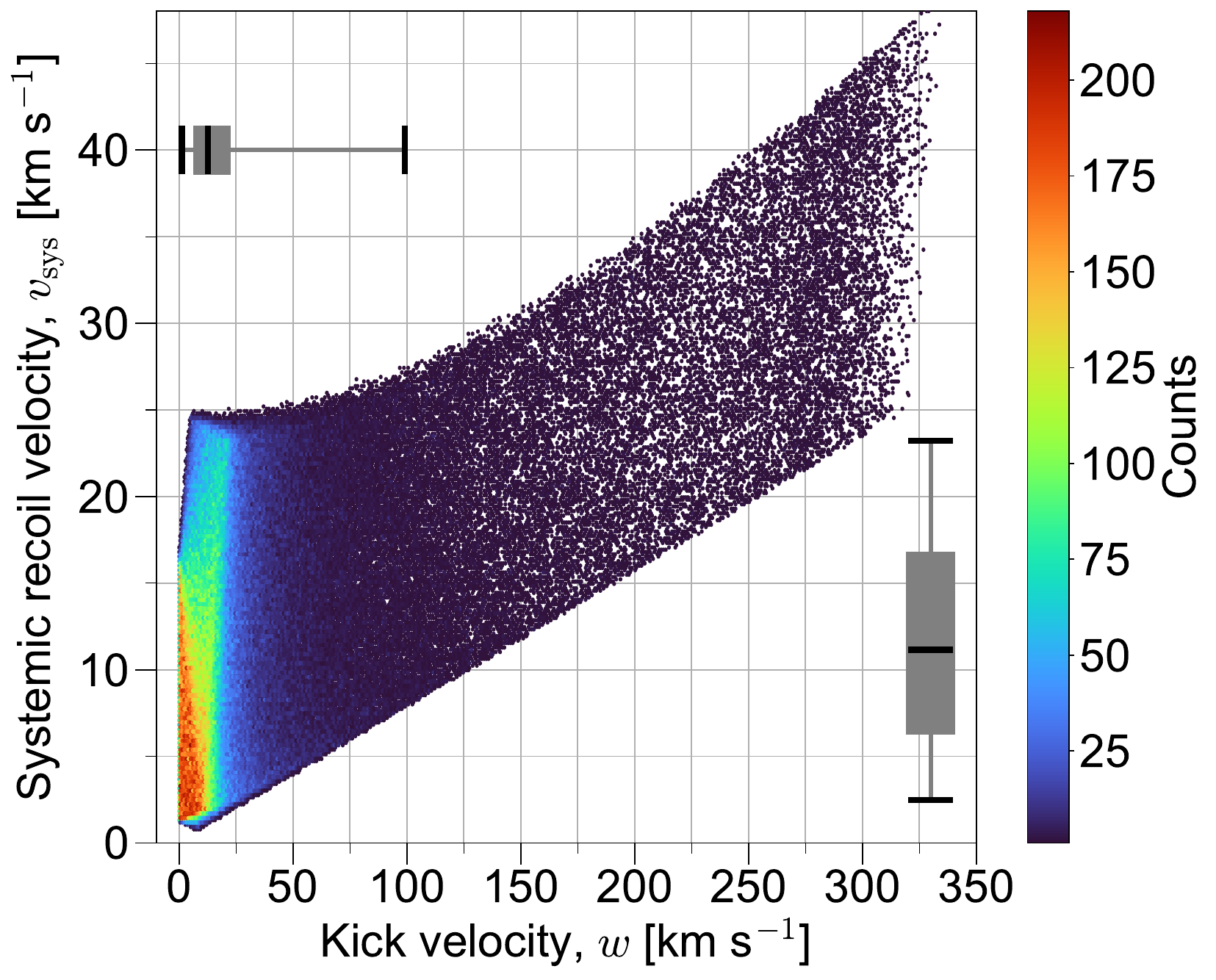}
\caption*{{\bf Extended Data Fig.\,3}\;\;
\textbf{Scatterplot of 3D systemic recoil velocity vs. kick velocity magnitude.}
The plot is based on our simulations of 500,000 systems that resemble SGR~0755$-$2933. The data pixels are color-coded according to their counts. The median and equal-tailed 50\% (grey) and 90\% (full range) credible intervals are plotted for each parameter. 
}
\end{figure*}


\begin{thebibliography}{47}
\expandafter\ifx\csname url\endcsname\relax
  \def\url#1{\texttt{#1}}\fi
\expandafter\ifx\csname urlprefix\endcsname\relax\def\urlprefix{URL }\fi
\providecommand{\bibinfo}[2]{#2}
\providecommand{\eprint}[2][]{\url{#2}}

\bibitem{rpe+23}
\bibinfo{author}{Richardson, N.~D.} \emph{et~al.}
\newblock \bibinfo{title}{A high-mass X-ray binary descended from an ultra-stripped supernova}.
\newblock \emph{\bibinfo{journal}{Nature}} 
\textbf{\bibinfo{volume}{614}}, \bibinfo{pages}{45} (\bibinfo{year}{2023}).

\bibitem{bdg+16}
\bibinfo{author}{Barthelemy, S.~D.} \emph{et~al.}
\newblock \bibinfo{title}{Swift detection of a likely new SGR: SGR~0755$-$2933}.
\newblock \emph{\bibinfo{journal}{ATel}} 
\textbf{\bibinfo{volume}{8831}}, (\bibinfo{year}{2016}).

\bibitem{dstj21}
\bibinfo{author}{Doroshenko, V.}, \bibinfo{author}{Santangelo, A.}, \bibinfo{author}{Tsygankov, S.~S.}, \bibinfo{author}{Ji, L.}
\newblock \bibinfo{title}{SGR~0755$-$2933: a new high-mass X-ray binary with the wrong name}.
\newblock \emph{\bibinfo{journal}{\aap}} 
\textbf{\bibinfo{volume}{647}}, \bibinfo{pages}{165} (\bibinfo{year}{2021}).

\bibitem{hil83}
\bibinfo{author}{Hills, J.~G.}
\newblock \bibinfo{title}{The effects of sudden mass loss and a random kick velocity produced in a supernova explosion on the dynamics of a binary star of arbitrary orbital eccentricity. Applications to X-ray binaries and to the binary pulsars}.
\newblock \emph{\bibinfo{journal}{\apj}} \textbf{\bibinfo{volume}{267}},
  \bibinfo{pages}{322} (\bibinfo{year}{1983}).

\bibitem{tkf+17}
\bibinfo{author}{Tauris, T.~M.} \emph{et~al.}
\newblock \bibinfo{title}{Formation of Double Neutron Star Systems}.
\newblock \emph{\bibinfo{journal}{\apj}}
  \textbf{\bibinfo{volume}{846}}, \bibinfo{pages}{170} (\bibinfo{year}{2017}).

\bibitem{tv23}
\bibinfo{author}{Tauris, T.~M.}, \bibinfo{author}{van den Heuvel, E.~P.~J.}
\newblock \emph{\bibinfo{title}{Physics of Binary Star Evolution. From Stars to X-ray Binaries and Gravitational Wave Sources}}.
\newblock \bibinfo{book}{Princeton University Press}
  (\bibinfo{year}{2023}).

\bibitem{tlp15}
\bibinfo{author}{Tauris, T.~M.}, \bibinfo{author}{Langer, N.}, \bibinfo{author}{Podsiadlowski, Ph.}
\newblock \bibinfo{title}{Ultra-stripped supernovae: progenitors and fate}.
\newblock \emph{\bibinfo{journal}{\mnras}}
  \textbf{\bibinfo{volume}{451}}, \bibinfo{pages}{2123} (\bibinfo{year}{2015}).
  
\bibitem{ywl10}
\bibinfo{author}{Yoon, S.-C.}, \bibinfo{author}{Woosley, S.~E.}, \bibinfo{author}{Langer, N.}
\newblock \bibinfo{title}{Type Ib/c Supernovae in Binary Systems. I. Evolution and Properties of the Progenitor Stars}.
\newblock \emph{\bibinfo{journal}{\apj}}
  \textbf{\bibinfo{volume}{725}}, \bibinfo{pages}{940} (\bibinfo{year}{2010}).

\bibitem{ljr+21}
\bibinfo{author}{Laplace, E.} \emph{et~al.}
\newblock \bibinfo{title}{Different to the core: The pre-supernova structures of massive single and binary-stripped stars}.
\newblock \emph{\bibinfo{journal}{\aap}}
  \textbf{\bibinfo{volume}{656}}, \bibinfo{pages}{58} (\bibinfo{year}{2021}).

\bibitem{dsm+13}
\bibinfo{author}{Drout, M.~R.} \emph{et~al.}
\newblock \bibinfo{title}{The Fast and Furious Decay of the Peculiar Type Ic Supernova 2005ek}.
\newblock \emph{\bibinfo{journal}{\apj}}
  \textbf{\bibinfo{volume}{774}}, \bibinfo{pages}{58} (\bibinfo{year}{2013}).

\bibitem{dko+18}
\bibinfo{author}{De, K.} \emph{et~al.}
\newblock \bibinfo{title}{A hot and fast ultra-stripped supernova that likely formed a compact neutron star binary}.
\newblock \emph{\bibinfo{journal}{Science}}
   \textbf{\bibinfo{volume}{362}}, \bibinfo{pages}{201} (\bibinfo{year}{2018}).

\bibitem{tlm+13}
\bibinfo{author}{Tauris, T.~M.} \emph{et~al.}
\newblock \bibinfo{title}{Ultra-stripped Type~Ic Supernovae from Close Binary Evolution}.
\newblock \emph{\bibinfo{journal}{\apjl}}
  \textbf{\bibinfo{volume}{778}}, \bibinfo{pages}{L23} (\bibinfo{year}{2013}).
  
\bibitem{mmt+17}
\bibinfo{author}{Moriya, T.~J.} \emph{et~al.}
\newblock \bibinfo{title}{Light-curve and spectral properties of ultrastripped core-collapse supernovae leading to binary neutron stars}.
\newblock \emph{\bibinfo{journal}{\mnras}}
  \textbf{\bibinfo{volume}{466}}, \bibinfo{pages}{2085} (\bibinfo{year}{2017}).

\bibitem{jtcf21}
\bibinfo{author}{Long, J.}, \bibinfo{author}{Tauris, T.~M.}, \bibinfo{author}{Chen, W.-C.}, \bibinfo{author}{Fuller, J.},
\newblock \bibinfo{title}{Novel Model of an Ultra-stripped Supernova Progenitor of a Double Neutron Star}.
\newblock \emph{\bibinfo{journal}{\apjl}}
  \textbf{\bibinfo{volume}{920}}, \bibinfo{pages}{L36} (\bibinfo{year}{2021}).
  
\bibitem{aaa+17c}
\bibinfo{author}{Abbott, B.~P.} \emph{et~al.}
\newblock \bibinfo{title}{GW170817: Observation of Gravitational Waves from a Binary Neutron Star Inspiral}.
\newblock \emph{\bibinfo{journal}{\prl}}
  \bibinfo{pages}{161101} (\bibinfo{year}{2017}).

\end{thebibliography}
\end{document}